# 3D Deep-learning-based Segmentation of Human Skin Sweat Glands and Their 3D Morphological Response to Temperature Variations

Shaoyu Pei, Renxiong Wu, Hao Zheng, Lang Qin, Shuaichen Lin, Yuxing Gan, Wenjing Huang, Zhixuan Wang, Mohan Qin, Yong Liu, Guangming Ni, *Member, IEEE*

*Abstract*—**Skin, the primary regulator of heat exchange, relies on sweat glands for thermoregulation. Alterations in sweat gland morphology play a crucial role in various pathological conditions and clinical diagnoses. Current methods for observing sweat gland morphology are limited by their two-dimensional, in vitro, and destructive nature, underscoring the urgent need for real-time, non-invasive, quantifiable technologies. We proposed a novel three-dimensional (3D) transformer-based multi-object segmentation framework, integrating a sliding window approach, joint spatial-channel attention mechanism, and architectural heterogeneity between shallow and deep layers. Our proposed network enables precise 3D sweat gland segmentation from skin volume data captured by optical coherence tomography (OCT). For the first time, subtle variations of sweat gland 3D morphology in response to temperature changes, have been visualized and quantified. Our approach establishes a benchmark for normal sweat gland morphology and provides a real-time, non-invasive tool for quantifying 3D structural parameters. This enables the study of individual variability and pathological changes in sweat gland structure, advancing dermatological research and clinical applications, including thermoregulation and bromhidrosis treatment.**

## I. INTRODUCTION

SKIN is the primary organ responsible for preventing the invasion of external bacteria [1] and regulating heat exchange between the body and the surrounding environment [2], with the sweat gland serving as the major cutaneous appendages crucial for thermoregulation [3]. As a skin structure exhibiting significant individual variability and pathological alterations, sweat glands demonstrate diverse morphological characteristics that reflect adaptive functional responses under different physiological and pathological conditions [4-6]. For instance, pathogenic glands in bromhidrosis are several times larger than their normal counterparts[7], while sweat gland atrophy or absence triggers hypohidrosis [6, 8]. Neutrophilic eccrine hidradenitis presents with edema and degeneration of eccrine sweat glands and ducts in early and chronic lesions, respectively [9]. Investigating the three-dimensional (3D) morphological characteristics of sweat glands offers substantial research potential. However, the observation of sweat gland morphology in medical practice has largely remained confined to skin biopsy [10], a two-dimensional, in vitro, and destructive experimental method that often yields untimely and inaccurate results. There exists a critical research gap for real-time, non-invasive, and quantifiable detection technologies.

Optical coherence tomography (OCT) has emerged as a promising non-invasive imaging technology, often referred to as "optical biopsy", capable of providing high-resolution, 3D imaging of biological tissues [11]. OCT can detect fine details of biological samples at depths of 1–2 mm below the surface and has been increasingly used for visualizing sweat glands [12]. Based on OCT's ability to capture detailed structural information, researchers have proposed several imaging processing methods [13] and neural network approaches [14-16] to extract sweat gland information from 3D OCT volume data. However, none of previous studies have analyzed or quantitatively assessed the 3D morphology of sweat glands, particularly their spiral structure. This gap highlights the need for a 3D segmentation approach and a quantitative assessment of their volume, surface area, and length, which are crucial for advancing the understanding of sweat gland functions，such as thermoregulation, and improving diagnostic strategies for related disorders.

Over the past few years, deep-learning-based automated medical image segmentation techniques have gained prominence in various benchmarks. CNN-based architectures, such as Res-Net [17], U-Net [18], Deep-Lab [19], Seg-Net [20], and their variants [20-26], have become the de facto standard for medical image segmentation. More recently, transformer-

Manuscript received XX XXXXXX 2025; revised XX XXXXX 2025; accepted XX XXXXXX 2025. Date of publication XX XXXXXX 2025; date of current version XX XXXXXX 2025. This work was supported in part by the National Science Foundation of China under Grants 61905036, China Postdoctoral Science Foundation under Grants 2021T140090 and 2019M663465, Fundamental Research Funds for the Central Universities (University of Electronic Science and Technology of China) Under Grant ZYGX2021J012, and Medico-Engineering Cooperation Funds from University of Electronic Science and Technology of China under Grant ZYGX2021YGCX019. (Corresponding author: Guangming Ni).

Shaoyu Pei, Renxiong Wu, Hao Zheng, Lang Qin, Shuaichen Lin, Yuxing Gan, Wenjing Huang, Zhixuan Wang, Mohan Qin, Yong Liu, Guangming Ni are with the School of Optoelectronic Science and Engineering, University of Electronic Science and Technology of China, Chengdu 611731, China (e-mail: guangmingni@uestc.edu.cn).

based models [27] have garnered significant attention, demonstrating superior performance in computer vision [28-30] and medical image analysis [31]. In prior work, Ding et al. proposed BCL-U Net, a modified U-Net that incorporates Long Short-Term Memory (LSTM) and hybrid dilated convolution for automatic extraction of fingerprints and sweat glands from OCT images [14]. While this method yielded accurate results, it was limited by its focus on 2D image properties and was unable to fully exploit the 3D characteristics of OCT images, limiting its capacity to segment the spiral-shaped sweat glands with continuous edges in 3D space. Another two noteworthy works that also focus on sweat glands are by Zhang et al., who proposed an improved 3D V-Net [32] with a dual-decoding structure (DDVN) to segment sweat glands and map fingerprints [15], and Zhou et al., who proposed a 3D CNN-based depth compression network (UDCNet) to achieve end-to-end 2D subcutaneous sweat pore extraction, addressing the dimensional limitation by considering the 3D nature of sweat glands [16]. However, the limited receptive field of this CNN-based approach restricts its ability to integrate global contextual information, which is crucial for accurately segmenting small sweat glands with fine details.

In light of these considerations, here we proposed a novel 3D transformer-based [27] heterogeneous model that learned long-range dependencies, extracted multi-scale and cross-channel features, and integrated both spatial and channel attention mechanisms to accurately segment the 3D structure of eccrine sweat glands in OCT-captured skin volumes. Our proposed network leveraged Swin-UNETR [33] as the backbone, incorporating Efficient Channel Attention (ECA) blocks [34] at the skip connections for the first time. The fusion of spatial and channel attention mechanisms significantly enhanced feature extraction, leading to improved segmentation performance. Additionally, we introduced Dynamic Large Kernel (DLK) [35] block and Dynamic Feature Fusion (DFF) [35] block, enabling the network to both enlarge the receptive field and reduce computational complexity. To the best of our knowledge, this work represents the first attempt to automatically and accurately segment the 3D spiral-shaped structure of sweat glands and quantify their 3D morphological characteristics. This approach addressed existing technological gaps by establishing a benchmark for normal 3D sweat gland morphology and quantitatively analyzed the 3D morphology of sweat glands under different temperatures. It also paved the way for further advancements in enhancing the understanding of sweat gland function (such as thermoregulation), studying its behavior in various physiological and pathological contexts, (such as bromhidrosis) and advancing dermatological research and clinical applications.

All 3D datasets, including the original 3D skin OCT images used, 3D segmentation results, and the corresponding ground truth can be downloaded and viewed from the link: https://tianchi.aliyun.com/notebook-zi/myDataSet#datalabId=195640

## II. METHODS

### A. Sliding Window-Based Long-Range and Contextual Method for Automated 3D Sweat Gland Segmentation

Sweat glands are coiled tubular structures located primarily within the epidermal and dermal layers of the skin, which can be characterized as small, slender targets. We opt for Swin UNETR [44] as the backbone, which employs a Swin Transformer encoder to address the limitations of traditional Transformer architectures [38] in image segmentation.

Swin Transformer encoder extracts feature representations at multiple resolutions using a shifted windowing mechanism for computing self-attention, allowing the network to effectively capture subtle local features by performing local self-attention within each window. For sweat gland segmentation with small and slender targets, choosing this encoder demonstrates remarkable performance. Meanwhile, as described in [30], the computational complexity of the global self-attention exhibits quadratic growth with image size, while the window-based self-attention makes the network more efficient for image segmentation tasks.

In this study, the Swin Transformer encoder uses a patch size of 2×2×2, with tokens projected into an embedding space of dimension 24 by the linear embedding layer, and all other network parameters are consistent with those in [33]. A fully convolutional neural network-based decoder fuses upsampled outputs with multi-resolution features from the five-stage encoder via skip connections.

### B. Joint Spatial-Channel Attention Mechanism

The shifted window multi-head self-attention in the encoder serves as a spatial attention mechanism, enabling our network to learn spatial relationships in the feature map, such as the shape and boundaries of sweat glands. However, this mechanism treats all channels equally, neglecting their distinct characteristics. To address this, we introduce a channel attention mechanism that adaptively recalibrates channel weights, emphasizing features like sweat gland textures to enhance segmentation performance.

While the squeeze-and-excitation (SE) block [36] is a common method for channel attention, its two fully connected layers for dimensionality reduction can hinder performance and efficiency. To overcome this, we employ the Efficient Channel Attention (ECA) block [34], which eliminates dimensionality reduction for more effective attention prediction.

### C. Architecture Heterogeneity between Layers

We propose a hybrid architecture for 3D sweat gland segmentation, dividing the decoder into shallow and deep layers to optimize feature extraction. In the shallow layers, where features are simple, we employ Res-blocks as channel concatenation, avoiding unnecessary complexity.

In the deep layers, the Dynamic Large Kernel (DLK) block and Dynamic Feature Fusion (DFF) module [35] are employed, making the most effective use of the large receptive field in the deep layer of the network to learn complex features and adaptively select the most important ones, enhancing the model's robustness and generalization ability.

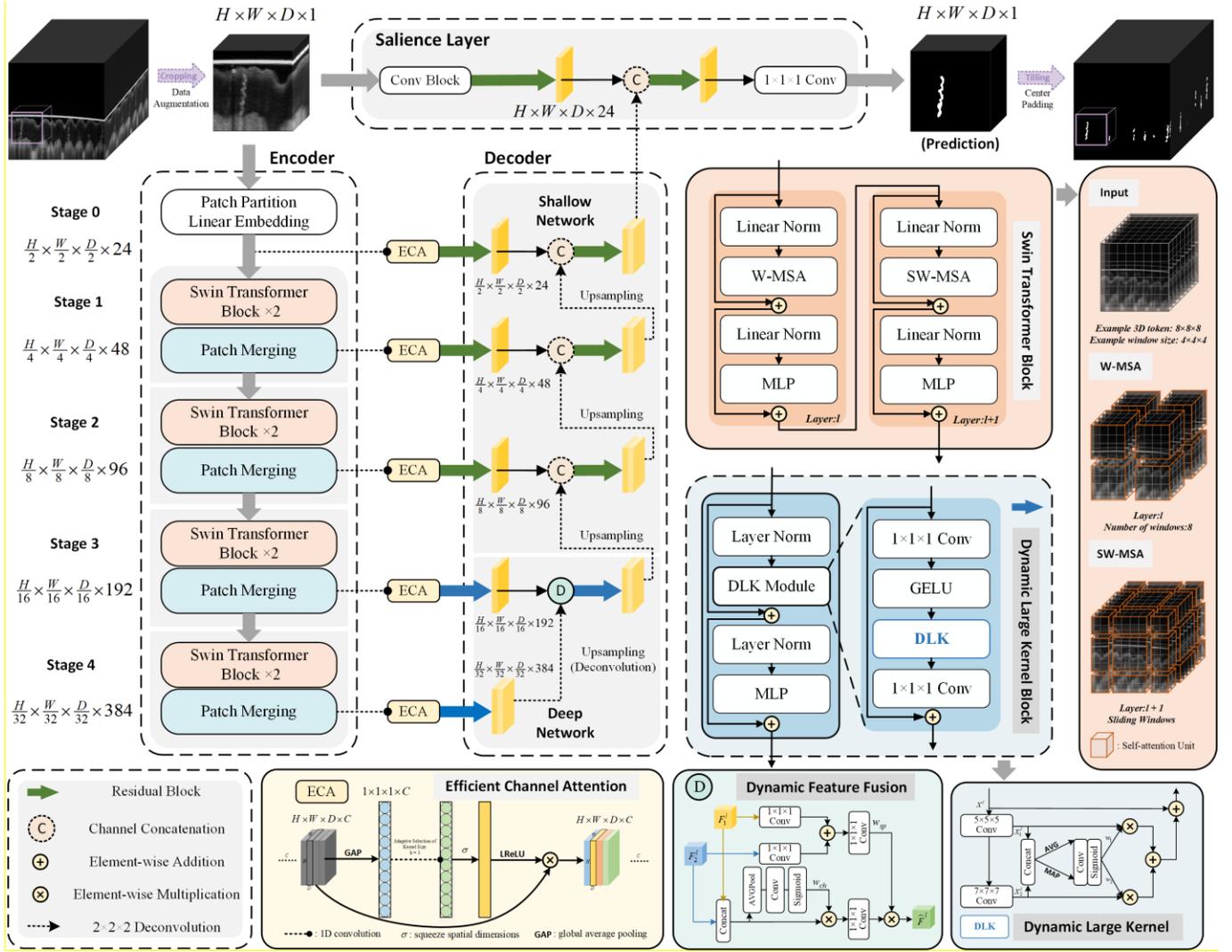

Fig. 1 Schematic overview of the proposed 3D segmentation model.

The DLK block uses progressively larger kernels with increasing dilation rates to expand the receptive field and adaptively emphasize critical features. The DFF module leverages global channel information to prioritize relevant feature maps and global spatial information to calibrate them, effectively highlighting salient regions.

In this study, the architecture and parameter configurations of the DLK and DFF blocks are consistent with those in [35].

### D. Hybrid Loss Function

The imbalanced class distribution poses major challenges for the sweat gland segmentation task, as the target pixels are significantly outnumbered by the background pixels, causing the model to favor generating enlarged foreground predictions. To address this issue, we adopt Binary Cross Entropy (BCE) [37] as the primary loss function. Additionally，to refine the segmentation boundaries and further improve precision, Dice loss [38] is incorporated. The hybrid loss function is:

$$L_{SweatGland} = \alpha L_{BCE} + \beta L_{Dice}. \quad (1)$$

The balance parameters are set as $\alpha = 0.9$ and $\beta = 0.1$.

### III. Experiments

#### A. Data Acquisition and Ground-Truth Label Generation

In the experiment, every skin OCT scan had an 800×800 scan pattern where a 3.9 mm×3.9 mm area was scanned with 800 horizontal lines (B-scans), each consisting of 800 A-lines per B-scan，resulting in an image size of 800×800×800 pixels (X×Z×Y). The axial and transverse optical resolutions are approximately 1.44 µm and 3.37 µm, respectively. In training, images were maintained at their original resolution to prevent potential distortions. In the results section, image sizes were rescaled to ensure the sweat gland morphology closely reflects its real structure. Our database contains a total of 150 skin OCT volume data from 40 different fingers in 10 different subjects, every finger was collected in three or four different regions. Despite the utilization of a flattening glass plate, shaking due to normal physiological reactions such as breathing and heartbeat are uncontrollable. Therefore, twelve OCT volume data containing 9600 B-scan images from different fingers were selected and manually annotated by dermatologists for better training.

Additionally, as Fig. 2 shows, *t*GT-OCT enhancement [39] was applied to enhance the raw OCT image quality. The size of the subblock after data augmentation is 288×288×64 pixels (X×Z×Y), and a total of 1567 subblocks are used for training.

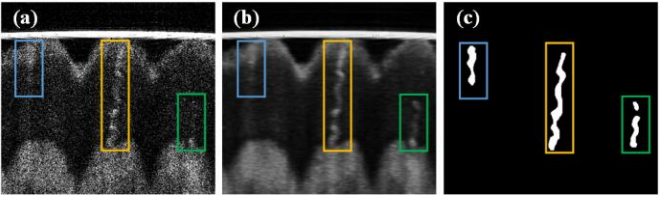

Fig. 2 Skin OCT image enhancement and annotation. Yellow boxes: glands visible before and after denoising; green boxes: glands clearer after denoising; blue boxes: glands unclear, requiring expert annotation. (a) Raw image, (b) Enhanced image, (c) Manual annotation.

### B. Experimental Implementation Details

Our method was trained with a batch size of 4, using AdamW optimizer with a weight decay of $1 \times 10^{-6}$ to prevent overfitting. The optimization process spanned 300 epochs, starting with a learning rate of $1 \times 10^{-5}$, followed by a 30-epoch warm-up to increase the learning rate gradually, then transitioning to cosine annealing for the remainder of the training. To accurately evaluate the proposed model performance, a 4-fold cross-validation strategy was employed, dividing the dataset into training and validation sets with a ratio of 3:1.

The models were implemented in Python v3.8.8, PyTorch v1.13.1, and MONAI v1.3.0, with data parallel training conducted across two nodes, each equipped with an NVIDIA RTX A6000 (48G) GPU.

## IV. RESULT

### A. Segmentation Results and Consistency Analysis

Fig. 3 displays the OCT skin 3D volume data (left), the corresponding sweat gland annotation (middle), and the 3D segmentation results (right), each accompanied by three B-scan images from different positions. Fig. 4 presents top-down views of the 3D segmentation outcomes from Fig. 3 and quantitative assessments of segmentation performance. Bland-Altman analysis indicates an average bias for volume measurements of $0.2981 \times 10^{-4}$ mm³, with 95% limits of agreement ranging from $-1.8897 \times 10^{-4}$ mm³ to $2.4859 \times 10^{-4}$ mm³ (Fig. 4(a)). For surface area measurements, the average bias was $-0.4281 \times 10^{-2}$ mm², with 95% limits of agreement spanning from $-3.4766 \times 10^{-2}$ mm² to $2.6205 \times 10^{-2}$ mm² (Fig. 4(b)). There is a strong correlation in sweat gland density measurements (Pearson: r = 0.982932, p < 0.00000001; Spearman: r = 0.956140, p < 0.000001) (Fig. 4(c)).

These results highlight the model's robust performance across volume, surface area, and density indicators, highlighting the method's potential for clinical applications.

### B. External Testing

Two OCT skin datasets were selected as the test set. The top-down views (Fig. 5(a4) and 5(b4)) provide an intuitive representation of sweat gland distribution, while B-scan images (Fig. 5(a1)-(a3) and 5(b1)-(b3)) highlight sweat gland predictions with semi-transparent red regions. Fig. 5(a5)-(a10) and 5(b5)-(b10) display local segmentation results, distinctly revealing the helical structure of the sweat glands.

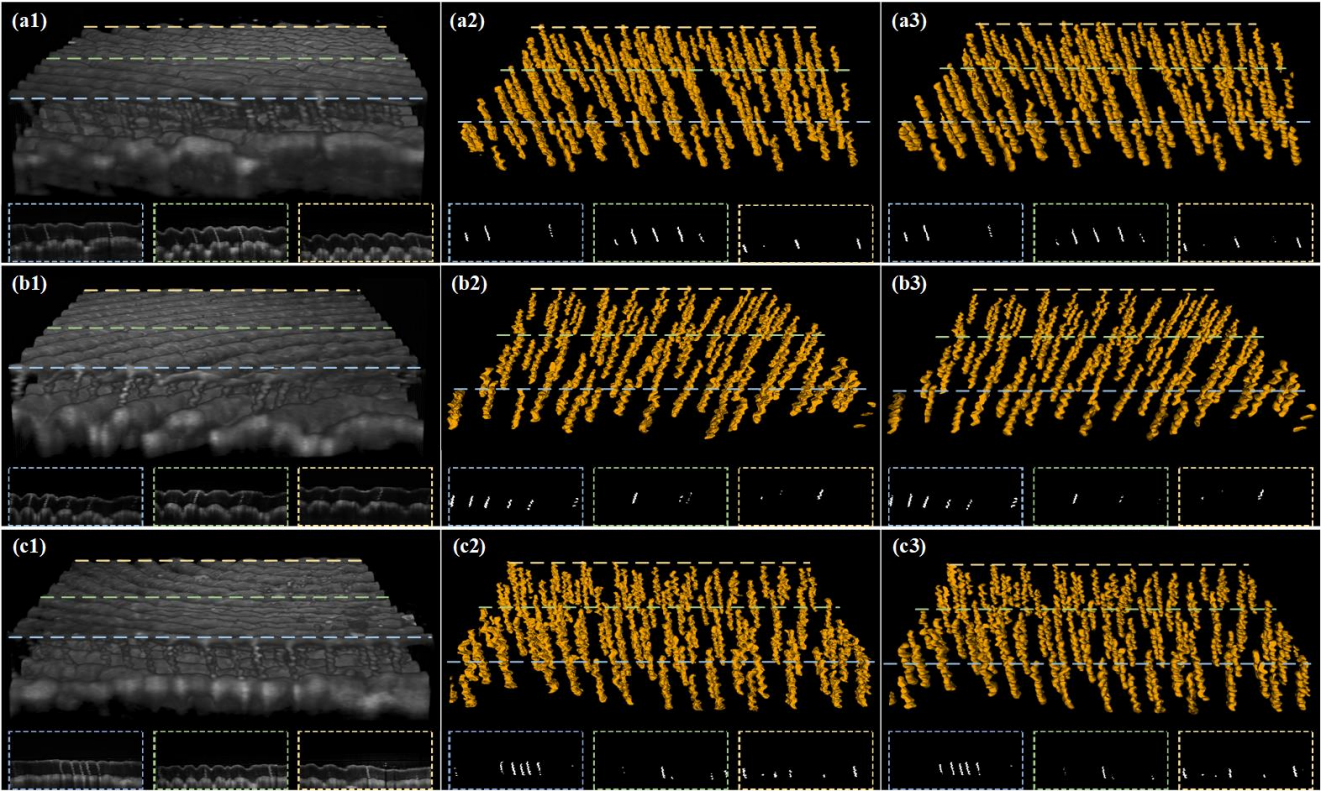

Fig. 3 3D segmentation results on the validation sets. Columns show the OCT-based 3D skin model ((a1)-(c1)), 3D annotation map ((a2)-(c2)), and 3D segmentation results ((a3)-(c3)). Blue, green, and yellow boxes indicate B-scan cross-sections at frames 1, 400, and 800, respectively.

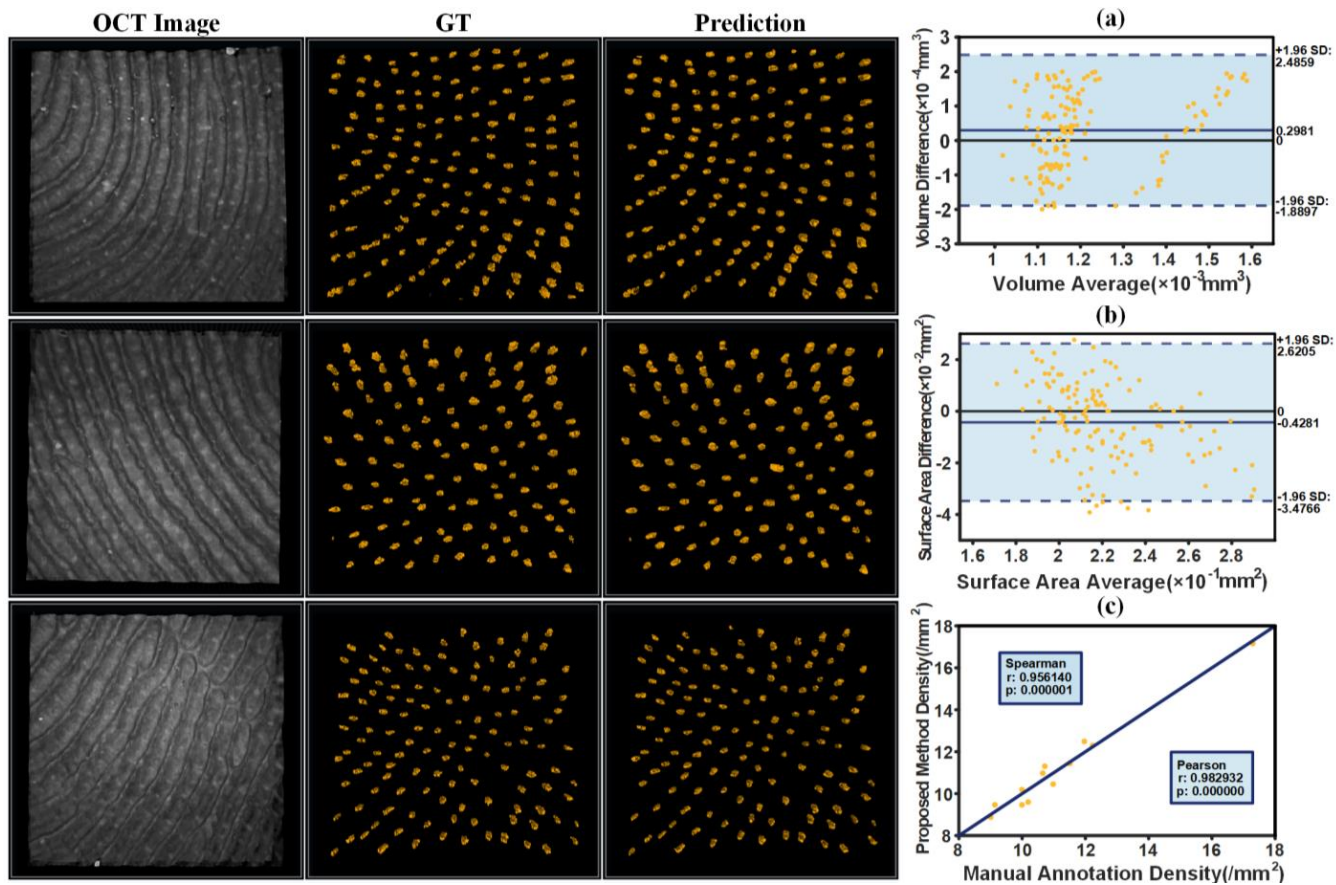

**Fig. 4** Top-down views (left panels). (a, b) Bland-Altman analyses for volume and surface area with bias (solid blue lines) and 95% limits of agreement (dashed lines) (c) Density measurements with Spearman and Pearson correlation analyses.

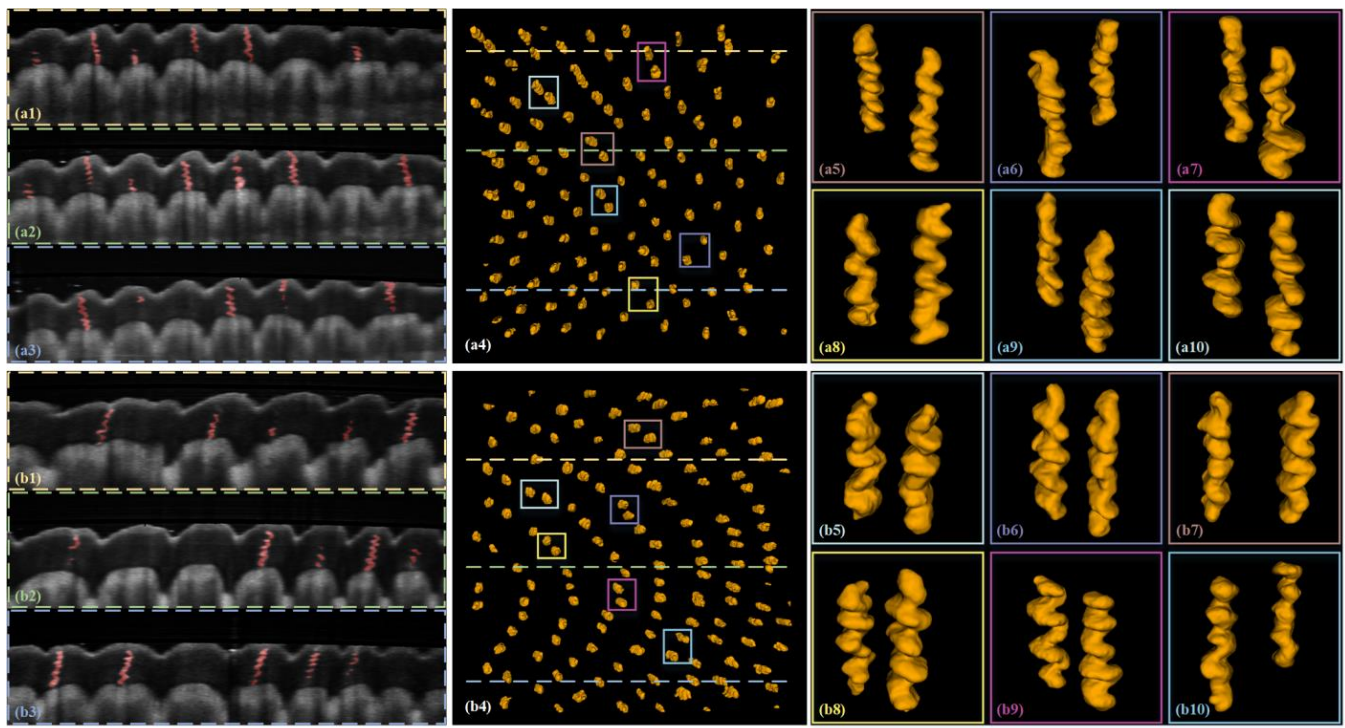

**Fig. 5** Top-down (a4, b4) and localized (a5–a10, b5–b10) 3D sweat gland segmentation for two finger skin regions. (a1–a3, b1–b3) show B-scans corresponding to dashed lines in (a4, b4).

This demonstrates the method's capability to effectively represent both global and local features, meeting the diverse demands of medical diagnosis with high precision.

## C. Evaluation and Comparisons

### 1) Limitations of Two-Dimensional Methods

In Fig. 6, we compare sweat gland boundary smoothness segmented by manual annotation, 2D U-Net [18], and the proposed 3D method using S/V ratio and local curvature metrics. As shown in Fig. 6(a), the 2D U-Net has the lowest S/V ratio ($0.12 \pm 0.007$) due to its inability to capture cross-slice spatial relationships, leading to incomplete gland morphology (blue arrows in Fig. 6(b)-(c)). Manual annotation shows a higher S/V ratio ($0.21 \pm 0.016$) than the proposed 3D method ($0.18 \pm 0.006$), as the annotation process preserves more boundary details but results in boundary inconsistencies. The blue bars in Fig. 6(a) depict the local curvature results, which show a decreasing trend from manual annotation ($0.0098 \pm 0.0018$) to the 2D method ($0.0085 \pm 0.0015$) and the proposed 3D method ($0.0080 \pm 0.0014$).

Our proposed 3D method holistically processes the sweat glands' 3D structure, yielding smoother, more continuous boundaries that closely match the true morphology.

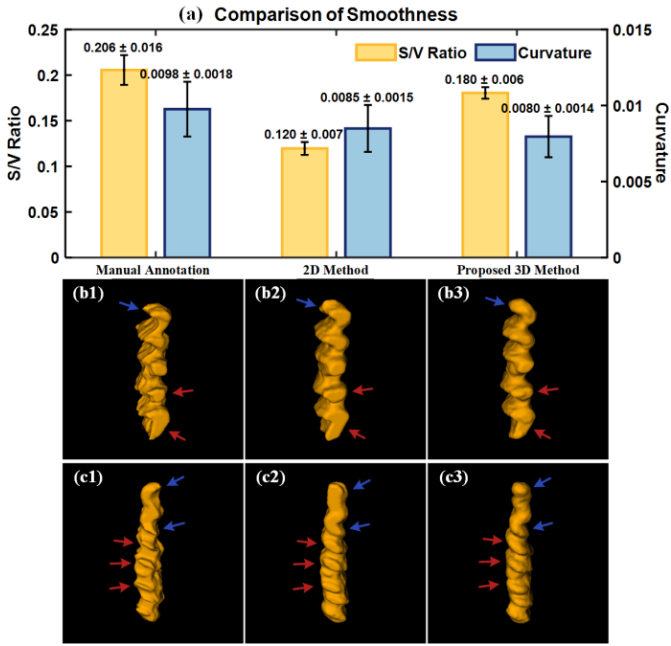

Fig. 6 Sweat gland boundary smoothness analysis. (a) S/V ratio and local curvature for Manual Annotation, 2D U-Net, and the proposed 3D method. (b1–b3) and (c1–c3) show the segmentation by each method. Blue arrows highlight features missed by 2D U-Net; red arrows indicate smoother, continuous morphology captured by the 3D method.

### 2) Comparison with Leading Three-Dimensional Methods

Based on Fig. 6, 2D segmentation is insufficient for accurate sweat gland morphology analysis. Consequently, we compared our proposed 3D approach with several state-of-the-art 3D methods, including CNN-based 3D-ResUnet and V-net [32], Transformer-based UNETR [40] and Swin UNETR [33] (baseline); and a self-adapting framework for medical image segmentation, nnU-Net [41] (3D cascade).

Our proposed 3D segmentation method achieved superior performance, establishing new state-of-the-art results across multiple metrics, including Dice ($0.8925 \pm 0.0355$), IoU ($0.8062 \pm 0.0450$), Precision ($0.8381 \pm 0.0692$), and Accuracy ($0.9955 \pm 0.0024$). As evident from the quantitative metrics in Table I, our method achieved a dual optimization of segmentation performance and model efficiency.

Detailed analysis of the comparative experiment are presented in Fig. 7: (a1)-(a7) show global segmentation accuracy, (b1)-(b14) focus on individual gland segmentations, and (c1)-(c14), (d1)-(d14) provide B-scan cross-sectional views. Blue and green boxes highlight the same gland across images, yellow arrows mark missed true positives, and red boxes and arrows denote false positives.

Our proposed method shows a slightly lower Recall ($0.9496 \pm 0.0912$) compared to 3D-ResUNet ($0.9938 \pm 0.0047$), V-Net ($0.9738 \pm 0.0137$), and UNETR ($0.9568 \pm 0.0483$), due to its focus on balancing Precision and Recall, which is crucial in reducing false positives and over-segmentation. As shown in Fig. 7, 3D-ResUNet, V-Net, and UNETR exhibit noticeable false positive regions.

Likewise, nnU-Net struggles to capture the intricate spiral structure of individual gland segmentations (Fig. 7(c5)-(c6)), making it less suitable for morphological studies of sweat glands. SwinUNETR network strikes a better balance between Recall and Precision, with fewer missegmentations, but under-segmentation is apparent.

Our proposed 3D method addresses these limitations by prioritizing channels with higher semantic relevance while suppressing less informative ones, effectively mitigating under-segmentation by strengthening the representation of key features. For small and complex targets like sweat glands, this improvement ensures more precise segmentation.

In conclusion, Fig. 7 and Table I show that the proposed method achieves accurate, consistent segmentation with reduced false positives and precise sweat gland boundary delineation. This highlights its robustness and potential for real-time morphological studies and quantitative sweat gland analysis.

TABLE I
EVALUATION RESULTS FOR SWEAT GLAND SEGMENTATION. STATE-OF-THE-ART 3D METHODS VERSUS OUR PROPOSED METHOD.

| Method | Dice | IoU | Precision | Recall | Accuracy | Total params |
|---|---|---|---|---|---|---|
| **Ours** | **0.8925±0.0355** | **0.8062±0.0450** | **0.8381±0.0692** | 0.9496±0.0912 | **0.9955±0.0024** | 7,375,884 |
| nnU-Net (3D-cascade) | 0.7802±0.0388 | 0.6472±0.0475 | 0.8195±0.0405 | 0.7518±0.0845 | 0.9923±0.0003 | 31,196,458 |
| Swin UNETR (baseline) | 0.8340±0.0401 | 0.7172±0.0554 | 0.8125±0.0762 | 0.8786±0.1124 | 0.9921±0.0020 | 15,505,249 |
| UNETR | 0.7975±0.0425 | 0.6648±0.0463 | 0.6853±0.0494 | 0.9568±0.0483 | 0.9898±0.0031 | 64,268,241 |
| V-Net | 0.7011±0.0422 | 0.5414±0.0494 | 0.5496±0.0515 | 0.9738±0.0137 | 0.9845±0.0053 | 4,051,954 |
| 3D-ResUnet | 0.6295±0.0801 | 0.4637±0.0753 | 0.4652±0.0759 | **0.9938±0.0047** | 0.9783±0.0062 | **1,185,818** |

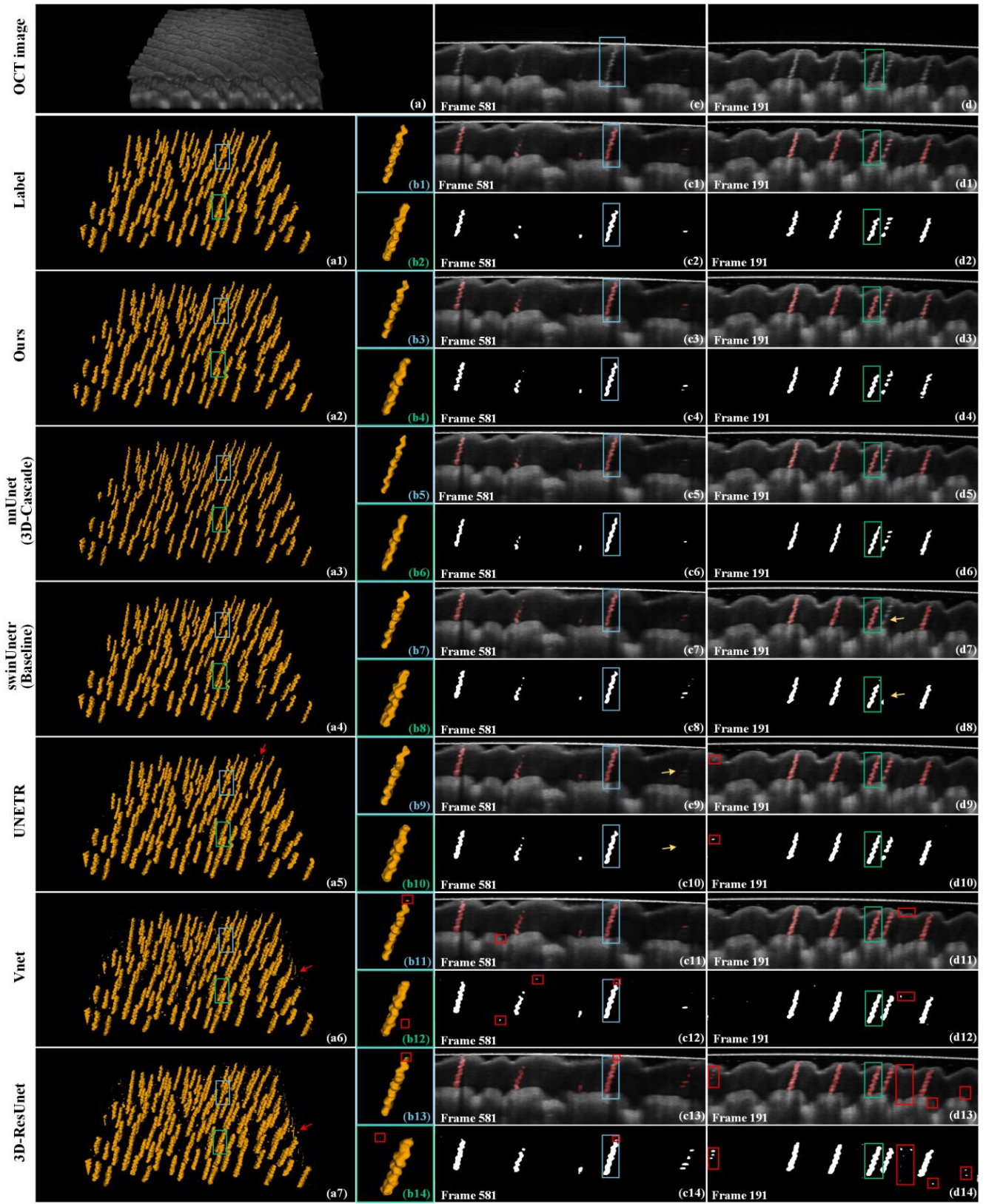

Fig. 7 Multidimensional visual comparisons between SOTA and our proposed methods. (a), (b), and (c) show 3D models of original OCT skin images with two B-scan slices at different locations. (a1)-(a7) present sweat gland segmentation results using manual annotations and various 3D methods, with blue and green boxes highlighting local segmentation results. (c1)-(c14) and (d1)-(d14) display B-scan images with annotations, showing under-segmented true positives (yellow arrows) and false positives (red boxes and arrows).

### D. Sweat Gland Morphological Parameter Quantification

To assess the clinical relevance of our method, we acquired OCT images of the same skin region under varying temperatures and applied our 3D model to segment sweat glands. Qualitative and quantitative analyses were then performed on the same glands across different conditions.

The study included 180 sweat gland samples from five subjects at 10°C, 33°C, and 43°C, with 60 samples per condition, ensuring a statistical power of 0.8 ($\alpha = 0.05$). Skin moisture was standardized across groups, and humidity differences were controlled. Each measurement was repeated six times per subject to minimize variability.

Fig. 8 displays segmentation results for 24 sweat glands, with eight subfigures ((a)-(h)) illustrating morphological changes under different temperatures. The first row shows 3D segmentation results, while the second row presents B-scan images at the glands' longest slices. As shown, with rising temperatures, most sweat glands demonstrate a significant increase in morphological parameters (Volume, Surface Area, Length). We combined quantitative and qualitative analyses to ensure reliability and generalizability, reducing subjectivity and enhancing precision. Table II and Fig. 9 present detailed quantitative results on sweat gland morphology, complementing the visual comparisons in Fig. 8.

Table II presents four sweat gland parameters across the three temperature groups. In univariate analysis, sweat glands at lower skin temperature (10°C) showed significantly reduced volume (10°C vs. 33°C: 2.5776±0.4617 vs. 3.2046±0.4034, $p = 2.56 \times 10^{-7}$), surface area (10°C vs. 33°C: 4.6610±0.4982 vs. 5.4207±0.4987, $p = 7.63 \times 10^{-8}$), and length (10°C vs. 33°C: 282.2169±20.5919 vs. 302.8383±20.3355, $p = 0.00015$). Conversely, at higher skin temperature (43°C), sweat glands exhibited significantly greater volume (43°C vs. 33°C: 3.6605±0.5395 vs. 3.2046±0.4034, $p = 0.00030$), surface area (43°C vs. 33°C: 5.8906±0.5139 vs. 5.4207±0.4987, $p = 0.00044$), and length (43°C vs. 33°C: 318.3632±15.4945 vs. 302.8383±20.3355, $p = 0.0011$).

Interestingly, the surface-to-volume ratio followed the opposite trend. Sweat glands at lower temperatures (10°C) had significantly higher surface-to-volume ratios compared to moderate temperatures (10°C vs. 33°C: 0.1834±0.01667 vs. 0.1701±0.01099, $p = 0.00037$), while at higher temperatures (43°C), the ratio was significantly lower (43°C vs. 33°C: 0.1624±0.01223 vs. 0.1701±0.01099, $p = 0.010$). Fig. 9 provides box-violin plots for an intuitive visualization of the data distribution recorded in Table II, offering a clear perspective on the variations in sweat gland parameters across the three temperature conditions.

This comprehensive analysis demonstrates our proposed method's ability to accurately segment 3D sweat glands and detect subtle morphological changes under different temperature conditions, providing valuable insights into sweat gland dynamics and their physiological responses to temperature variations.

Previous studies using histological sectioning focused on long-term temperature-induced sweat gland changes [2, 6]. In contrast, our 3D analysis based on real-time OCT imaging reveals significant 3D morphological adaptations even under short-term temperature variations. At lower temperatures (10°C), sweat glands exhibit reduced activity, with decreased volume, surface area, and length, leading to a "constricted" state. At higher temperatures (43°C), glands become highly active, showing increased size and a "dilated" morphology.

In addition, we identified an unexpected finding: the surface-to-volume ratio (S/V ratio) of sweat glands at higher temperatures was significantly lower than at moderate (33°C) or lower temperatures. A plausible explanation is that sweat gland morphology prioritizes functional adaptation under varying thermal conditions. At higher temperatures, sweat glands may preferentially increase their lumen volume to enhance sweating efficiency while minimizing energy loss associated with excessive surface area. Conversely, at lower temperatures, maintaining a higher S/V ratio could optimize metabolic exchange between the sweat glands and surrounding tissues under reduced activity levels.

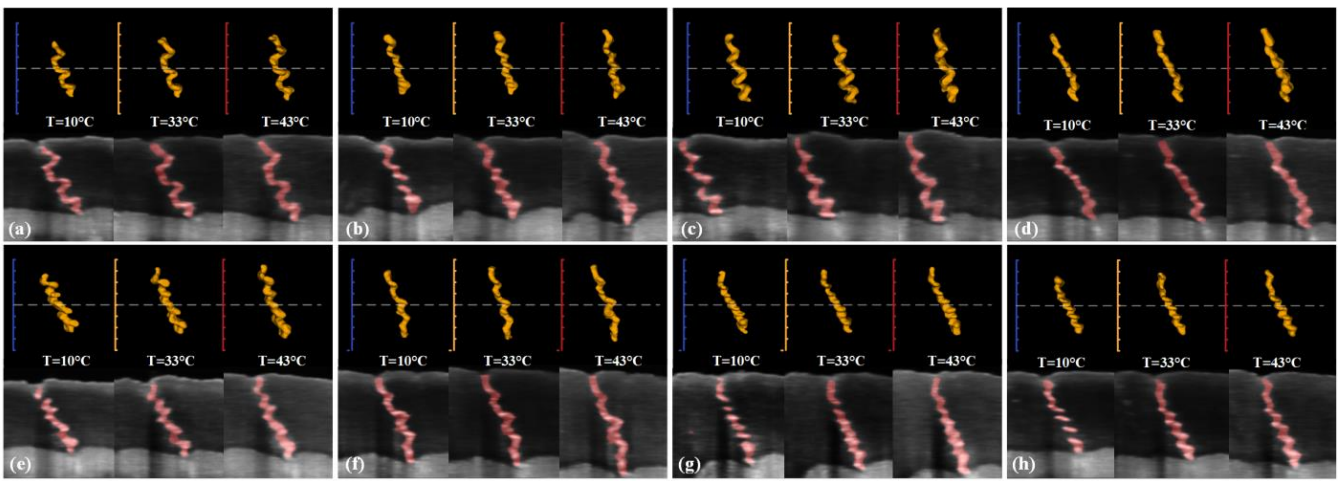

Fig. 8 Morphological changes of eight sweat glands at 10°C, 33°C, and 43°C. Each subfigure ((a)–(h)) shows 3D segmentation models (top) and corresponding B-scan results (bottom).

QUANTITATIVE ANALYSIS OF 3D MORPHOLOGICAL METRICS OF SWEAT GLANDS UNDER DIFFERENT TEMPERATURES (T = 10°C, 33°C, 43°C).

| Sweat Gland Parameters | T = 10°C (n=5) | T = 33°C (n=5) | T = 43°C (n=5) | p-value (10°C vs. 33°C) | p-value (43°C vs. 33°C) | p-value (10°C vs. 43°C) |
|---|---|---|---|---|---|---|
| Volume ($\times 10^{-4} mm^3$) | 2.5776±0.4617 | 3.2046±0.4034 | 3.6605±0.5395 | $2.56 \times 10^{-7}$ | 0.00030 | $3.2614 \times 10^{-12}$ |
| Surface Area ($\times 10^{-2} mm^2$) | 4.6610±0.4982 | 5.4207±0.4987 | 5.8906±0.5139 | $7.63 \times 10^{-8}$ | 0.00044 | $4.4451 \times 10^{-14}$ |
| Length ($\mu m$) | 282.2169±20.5919 | 302.8383±20.3355 | 318.3632±15.4945 | 0.00015 | 0.0011 | $5.1402 \times 10^{-11}$ |
| S/V Ratio | 0.1834±0.01667 | 0.1701±0.01099 | 0.1624±0.01223 | 0.00037 | 0.010 | $3.0474 \times 10^{-7}$ |

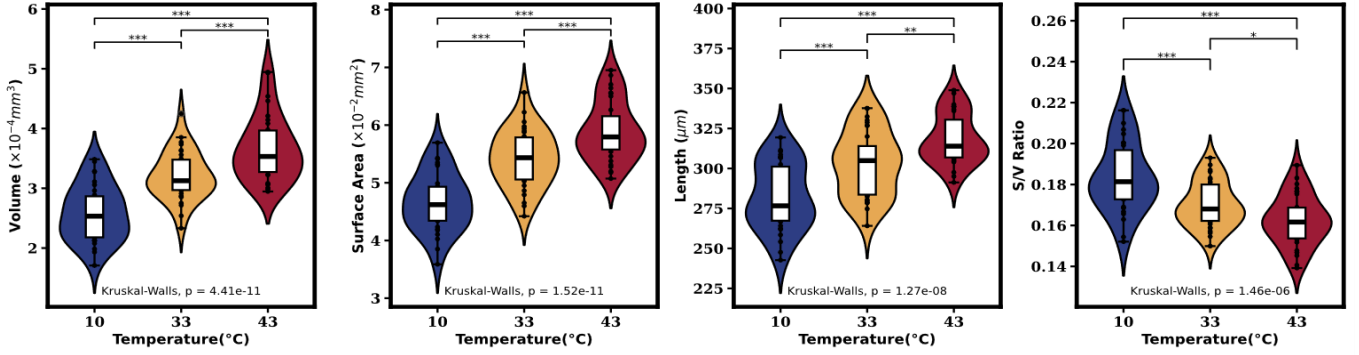

Fig. 9 Box-violin plots of 3D sweat gland metrics under different temperatures (T = 10°C, 33°C, 43°C) with statistical significance (∗$p<0.05$, ∗∗$p<0.01$, ∗∗∗$p<0.001$) based on the Kruskal-Wallis test.

These findings offer novel insights into the morphological and functional adaptations of sweat glands to varying thermal environments, shedding light on the thermoregulatory roles of sweat glands in human skin and providing a fresh perspective on their physiological behavior.

## V. DISCUSSION

We propose a transformer-based 3D neural network for sweat gland 3D segmentation in OCT skin images, precisely capturing their three-dimensional helical morphology. This approach addresses the research gap in previous deep-learning methods, which have not analyzed or quantitatively assessed the 3D morphology of sweat glands, particularly their spiral structure. After validating the accuracy of 3D sweat gland segmentation, we applied our model to predict sweat glands in OCT images of the same skin region under different temperatures, demonstrating the clinical applicability of our proposed method.

In prior studies on sweat gland morphology, methodological limitations hindered real-time quantitative analyses of sweat gland structures. The proposed method enables non-invasive, real-time, and precise delineation of 3D spiral-shaped sweat glands, providing a new reference for assessing normal sweat gland morphology and facilitating quantitative comparisons of morphological parameters. Notably, we are the first to quantitatively demonstrate adaptive morphological changes in sweat glands in response to short-term environmental temperature variations.

These breakthroughs establish a benchmark for normal sweat gland morphology and provide quantitative comparative analyses of morphological parameters, demonstrating the method's clinical significance in dermatological practice.

Future work will focus on reducing the reliance on labor-intensive 2D manual annotations by exploring semi-supervised or unsupervised frameworks, as well as developing strategies for direct 3D annotation, which remains a major challenge due to the complexity of volumetric data labeling. Additionally, further investigation is needed to elucidate the observed decrease in the surface-to-volume ratio with rising temperatures.

To better explore individual variability and pathological changes in sweat gland morphology, future studies should optimize OCT device configurations to capture data from a broader range of regions, such as the armpits, face, palms, and soles, as well as from sweat glands exhibiting pathological alterations. Integrating these datasets with our proposed 3D segmentation method will enable more comprehensive and diverse qualitative and quantitative analyses of sweat gland morphology.

## VI. CONCLUSION

Here we propose a novel 3D deep learning-based heterogeneous model that integrates long-range dependency modeling, multi-scale feature extraction, and spatial-channel attention mechanisms. This model enables precise segmentation of the 3D spiral morphology of sweat glands in OCT-captured skin images and quantifies subtle 3D morphological changes, establishing a benchmark for normal sweat gland morphology while providing a real-time, non-invasive tool for clinical analysis.

By bridging current technological gaps, the integration of this 3D segmentation framework with OCT technology expands the potential for studying sweat gland function (such as thermoregulation) and 3D morphological behavior of sweat glands across a wide range of physiological and pathological conditions, offering deeper insights into the adaptive mechanisms of the skin (such as bromhidrosis).